\documentclass{article}
\usepackage{ijcai20}

\usepackage{times}
\usepackage{soul}
\usepackage{enumitem}
\usepackage[table]{xcolor}
\usepackage{url}
\usepackage[hidelinks]{hyperref}
\usepackage[utf8]{inputenc}
\usepackage[small]{caption}
\usepackage{graphicx}
\usepackage{amsmath}
\usepackage{amsthm}
\usepackage{booktabs}
\usepackage{algorithm}
\usepackage{algorithmic}
\title{On Fairness and Interpretability}

\author{
     Deepak P$^1$ \hspace{0.2in} Sanil V$^2$ \hspace{0.2in} Joemon M. Jose$^3$
     \affiliations
     $^1$Queen's University Belfast \hspace{0.2in} $^2$IIT Delhi \hspace{0.2in} $^3$University of Glasgow
     \emails
     deepaksp@acm.org \hspace{0.2in} sanil@hss.iitd.ernet.in \hspace{0.2in} Joemon.Jose@glasgow.ac.uk 
}

\begin{document}

\maketitle

\begin{abstract}
Ethical AI spans a gamut of considerations. Among these, the most popular ones, fairness and interpretability, have remained largely distinct in technical pursuits. We discuss and elucidate the differences between fairness and interpretability across a variety of dimensions. Further, we develop two principles-based frameworks towards developing ethical AI for the future that embrace aspects of both fairness and interpretability. First, {\it interpretability for fairness} proposes instantiating interpretability within the realm of fairness to develop a new breed of ethical AI. Second, {\it fairness and interpretability} initiates deliberations on bringing the best aspects of both together. We hope that these two frameworks will contribute to intensifying scholarly discussions on new frontiers of ethical AI that brings together fairness and interpretability.
\end{abstract}

\section{Introduction}

AI algorithms are being increasingly used for decision making within scenarios with social and political consequences (e.g., benefit eligibility, recidivism) as opposed to traditional automation scenarios (e.g., handwriting recognition). This has rightly spurred recent interest in {\it Ethical AI}. The broad umbrella of ethical AI or responsible AI~\cite{dignum2019responsible} involves considerations such as fairness~\cite{chouldechova2020snapshot}, interpretability\footnote{Interpretability, as we use in this paper, is quite related to, and often interchangeable with the notion of explainability.}~\cite{molnar2020interpretable}, privacy~\cite{mohassel2017secureml} and trustworthiness~\cite{toreini2020relationship}. Of these, fairness and interpretability are arguably the two considerations that have been explored quite heavily. Certain pairs of considerations, such as interpretability and trustworthiness, may be seen as apparently synergistic. There is much less understanding of how technological efforts across other pairs of considerations, such as fairness and interpretability as we consider here, can work together.

\noindent{\bf Our Contributions:} In this position paper, we first discuss the conceptual distinctions between fairness and interpretability as it applies to AI/ML. Next, we propose two frontiers of AI research in which efforts towards fairness and interpretability can be meaningfully blended towards advancing ethical AI in unique and novel ways. Where appropriate, we will use example scenarios from unsupervised data-driven AI to illustrate the arguments. This is motivated partly by the fact that unsupervised learning is relatively less explored within ethical AI, making it malleable to fresh thought leads. 


\section{Fairness \& Interpretability: Distinctions}

Fairness and Interpretability have largely been, within scholarly literature, seen as two distinct and different frontiers under the broader {\it Ethical AI} umbrella. The distinctions may be characterized under various dimensions as we discuss below. 

{\it First}, the family of fairness principles~\cite{narayanan2018translation} are normative values that relate to the politics of AI-driven decision making~\cite{wong2019democratizing}. On the other hand, interpretability considers user's ability to understand decisions, and lies at the interface between humans and AI. In other words, technological efforts towards deepening fairness would benefit from better grounding on political philosophy, whereas interpretability would have an analogous relationship with behavioral sciences. In fact, much interpretable and explainable AI work has appeared within HCI-related computing avenues. {\it Second}, there is a tension between fairness and accuracy (or any other utilitarian metric, say, efficiency), and similarly, there is a tension between interpretability and accuracy. However, these tensions are in {\it different directions}, as recently outlined in~\cite{kleinberg2019simplicity}. The authors illustrate that a simple and interpretable model can be strictly improved on both fairness and accuracy by making it more complex (thus reducing interpretability). In other words, there exists a tension between fairness and interpretability. {\it Third}, a system that produces interpretable results enhances user's trust in the AI in ways that a fair AI may not (at least, in the short-term). This means that interpretability is arguably likely to be more {\it 'popular'}, and thus would be prioritized over fairness by the private sector which is where most AI is developed. {\it Fourth}, interpretability can be assessed at the level of individual decisions made by the AI, whereas fairness assessments need to go much beyond analyzing individual decisions. Fairness assessment often involves a normative evaluation of the process and/or the distribution of decisions made. {\it Fifth}, there is a distinct contrast between the subjectivity of fairness and interpretability. Fairness is subjective at a normative level, and the subjectivity is often grounded in political positions; supporters of {\it individual fairness} are likely to be to the right of {\it group fairness}, within the left-right political spectrum. On the other hand, interpretability is often highly subjective in {\it politically neutral} ways; the same result or explanation may be regarded as less interpretable by one user, whereas it could be rated as more interpretable by another. This implies that interpretability is amenable to personalization (potentially through data-driven A/B testing, when user feedback is available) at a much more extensive level than fairness is. {\it Lastly}, it may be argued that different application domains of AI have different priorities between interpretability and fairness. Interpretability may be considered very important in fields involving high-bandwidth human-technology interaction such as robotics and HCI. On the other hand, fairness may be of prime importance in societally relevant applications such as automation of applicant screening for jobs, policing and automated decision making on benefit and healthcare eligibility. 

The distinctions discussed above do not just apply to fairness vs. interpretability. While other theories from ethical philosophy share many features with fairness as discussed above, dimensions such as privacy, trustworthiness and transparency share several characteristics with interpretability. 

\section{Frontiers of Synergy}

We develop two frontiers of synergy for technical efforts towards interpretability and fairness. Specifically, the directions we outline require that technological building blocks situated within either of the two (interpretability and fairness) work together to achieve meaningful advances within ethical AI. We neither target nor accomplish a conceptual unification of the concepts of interpretability and fairness. 


\begin{table}[h]
\begin{center}
\begin{tabular}{c} 
\cellcolor{gray!25}{\it 'Justice must not only be done, but must be seen to be}\\
\cellcolor{gray!25} {\it done'} - Chief Justice Hewart (UK High Court, 1924) \\
\end{tabular}
\end{center}
\vspace{-0.2in}
\end{table} 

\section{Interpretability for Fairness}

\noindent{\bf Motivation:} Consider a decision that is output by a system that is purportedly {\it 'fair'}. How can somebody at the receiving end of a decision from such a system be sure that it is fair, according to the notion of fairness used in the system? This question may be instantiated based on the specific notion of fairness used, as follows. For a system that claims to use Rawlsian fairness, how can we ascertain that the present outcome is reasonable to ensure that the system functions in a way that is most advantageous to the least diadvantaged (Ref. {\it difference principle}~\cite{rawls1971theory}). Or in the case of {\it demographic parity} as fairness, how can one be sure that the decision is a natural fallout of a process that is designed to achieve parity across sensitive groups. Particularly, users are likely to ask these questions when they find themselves at the receiving end of what they perceive as a {\it bad} decision (e.g., denial of welfare support). In the best interests of explainability and in making sure that {\it fairness is perceived to be ensured} (alluding to the 1924 quote above), we (i) ought to answer these {\it 'how is this fair?'} questions, and (ii) do so without reference to other decisions for other specific individuals (for privacy). 

\noindent{\bf Background:} Having motivated that some sort of fairness explanations are necessary, we now consider ethical theories in the space. {\it Accountability for Reasonableness (AFR)}~\cite{daniels2008accountability} is an ethical framework designed for healthcare scenarios, in particular, when fairness is to be accounted for in scenarios involving allocation of scarce healthcare resources. Of particular interest to us is AFR's {\it relevance} condition that suggests that decisions are explained by appealing to rationales that are reasonable enough to be accepted by {\it fair-minded} people who are disposed to finding justifiable terms of co-operation. Badano~\cite{badano2018if} generalizes this to require acceptance by {\it each} reasonable person (aka {\it full acceptability}), which implicitly requires that those subject to most adverse decisions also be convinced. Recently, Wong~\cite{wong2019democratizing} has argued that AFR could provide directions towards addressing the political dimensions of algorithmic fairness. 

\noindent{\bf Interpretability for Fairness:} We propose a novel framework, {\it Interpretability for Fairness} (IFF), drawing inspiration from AFR. IFF blends AFR with the {\it design for values} (Ref.~\cite{dignum2019responsible} Sec 4.4) approach to formulate a set of principles targeted at using interpretability as a pathway to enhance acceptability of fair AI. The two IFF principles are:

\begin{itemize}[leftmargin=*]
    \item {\it Fairness Publicity Condition:} The fairness value(s) that are sought to be achieved by the AI system must be laid out clearly in layman-friendly language as comprehensively as possible. If a trade-off between values is intended (as often sought, such as a balance between utilitarianism and demographic parity fairness), the relative importance between the values in the mix should be exemplified. 
    \item {\it Values to Decision Interpretability Condition:} The system should strive to produce a layman-friendly and simple interpretation of each decision substantiating how it relates to the mix of values embodied in the system, as outlined in the publicity condition. This interpretation should be acceptable to any reasonable person who is disposed to finding mutually justifiable terms of co-operation. 
\end{itemize}

\noindent These principles are intended to be meaningful to a data scientist equipped with a reasonable understanding of the nuances and social aspects of the domain in which the AI is being designed to operate in, as opposed to abstract ones that pose a {\it 'translation'} challenge. In particular, IFF is at a lower level of abstraction due to instantiating accountability as interpretability and reasonableness as fairness-oriented reasonableness. IFF also keeps process governance aspects (e.g., appeals as in AFR) out of its scope and is focused on the technical design of the AI. While the first IFF condition draws from the analogous condition in AFR, the second condition is inspired by the {\it 'design for values'}~\cite{aldewereld2015design} maxim on linking values to concrete software functionalities. The second condition requires {\it fairness interpretability}, or explaining a decision based on the fairness values and any trade-offs with non-fairness values used in the system. We note here that the layman-friendliness requirement in the first condition entails exclusion of socio-technical terms such as {\it structural discrimination}~\cite{pincus1999individual} and {\it intersectionality}~\cite{carastathis2014concept}, whereas that in the second condition would require exclusion of AI-terminology such as {\it bayesian inference}~\cite{ghosal2017fundamentals}. 

\noindent{\bf IFF and Current Solutions:} IFF requires interpretability that is distinctly different from interpretability in the sense it is used in current Fair ML literature. Within our focus area of unsupervised learning, interpretability is dominated by rule-learning where features for rules are drawn from the data~\cite{balachandran2012interpretable} or auxiliary features~\cite{sambaturu2020efficient} such as tags. While these enable human-understandable descriptions of the outcomes, they do not satisfy the IFF requirement of explaining the fairness. Recent work on fair representation learning~\cite{he2020geometric} has posited that fairness-based re-engineered versions of original features may be considered interpretable as long as they remain attached to the semantic labels (e.g., {\it maths marks}, {\it annual income}) associated with features. However, IFF requires that the {\it re-engineering process be interpretable on the basis of fairness values}; for example, if the re-engineering transforms annual income for an individual from {\it \$40k} to {\it \$35k} prior to further downstream processing, IFF requires that this transformation be explained as a fallout of a reasonable fairness-seeking process. We haven't come across Fair ML work that may be argued to be conformant to IFF. 

\noindent{\bf IFF-aligned AI:} IFF is construed, much like AFR, as a set of guiding principles, and is not prescriptive as to {\it how} conformance may be achieved. By way of an illustrative example, a feature re-engineering method that {\it corrects the dimensions of achievement of socially discriminated demographics (e.g., backward castes) upward} may be acceptable by reasonable people as a fallout of a process targeting demographic parity, as long as the re-engineering process is interpretable. This is similar in spirit to differentiated age and attempt limits as enforced in India's affirmative action process\footnote{\scriptsize e.g., \url{https://en.wikipedia.org/wiki/Civil\_Services\_Examination_(India)\#Age}}. 

\noindent{\bf Why IFF?} As indicated in the motivation, IFF has the potential to deepen user confidence and trust in fairness-seeking algorithms through explicitly illustrating the conformance to fairness. A detailed treatment of the various aspects of the two IFF principles is not feasible due to space constraints.


\section{Fairness and Interpretations}

\noindent{\bf Motivation:} Consider using fair AI {\it along with} state-of-the-art approaches for interpretability that explain the outcomes using data or auxiliary information (not the IFF fairness interpretability). The absence of a connection between the two may lead to a dissonance between their outputs, especially for fair AI that operationalizes notions of group fairness. The fairness-agnostic search for user-friendly explanations could lead to manifestly unimpressive explanations. For example, the membership of an individual within a group could be motivated by demographic fairness considerations, but an explanation referring to a sensitive attribute such as ethnicity may be unacceptable, since that may be perceived as demeaning individuality. On the other hand, excluding sensitive attributes from the realm of explanations may lead to lower quality (thus, unacceptable) explanations. Either of the above could undermine user's trust in the AI, given recent research linking trust and explanation quality~\cite{kunkel2019let}. 

\noindent{\bf Fairness and Interpretations:} Towards addressing the above conundrum, we propose a layered paradigm, called {\it Fairness and Interpretations} (F\&I). {\it First}, we propose that an AI system be constrained to conform to  \underline{both}: (i) fairness, in accordance with the fairness values it targets, and (ii) reasonable interpretability, i.e., be able to provide reasonable explanations for its decisions. This rules out fair AI that is not reasonably interpretable, and vice versa. There could be several ways of characterizing reasonableness in explanations. One way would be to characterize reasonableness as individual fairness; in other words, the space of reasonable explanations may be characterized as being a space where the associated outcomes are {\it smooth}. Others ways could be to choose explanation paradigms (of which there are many~\cite{binns2018s}) that enhance user's perception of system fairness (an empirical study appears at~\cite{dodge2019explaining}). {\it Second}, for scenarios where both fairness and reasonable interpretability are hard to satisfy together (as could be the case where the data has high degrees of bias), we propose that the user be informed that no interpretable explanation can be supplied for the decision, and that the result be accompanied by {\it fairness explanations} as outlined in IFF. Thus, an F\&I-compliant AI is one that adheres to {\it fairness} (as designed for), and provides either {\it reasonable decision interpretability} or {\it only fairness explanations}. Additionally, the two F\&I principles are ordered lexically; unless there is a good reason that the first principle cannot be satisfied, the second does not come into play. The appeals process for such a system needs to be cognizant of whether the decision was accompanied by {\it reasonable explanations} or just {\it fairness explanations}; a higher appeal rate for the latter may be expected and planned for.

\noindent{\bf Why F\&I?} F\&I is intended as a paradigm that will meaningfully bring both {\it fairness} and {\it interpretability} together without artifacts of one dampening the other. While a detailed analysis of F\&I consequences is infeasible here due to space, F\&I-aligned AI is expected to be able to fuse normative and user-oriented aspects towards enhancing ethical AI. F\&I is well-aligned with and intends to further operationalization of the {\it right to explanations} enshrined within  GDPR~\cite{selbst2018meaningful} recommendations of the European Union. 

\noindent{\bf Technical Challenges:} In contrast to IFF, F\&I lies in a technically pristine space and entails crisp technical challenges. First, the notion of {\it reasonable explanations} needs to be technically instantiated, and computational approaches to determine reasonableness effectively and efficiently needs to be developed. Second, the twin constraints (fairness and reasonable explanations) would need to be achieved together, requiring novel multi-criteria optimization methods. Third, a decision procedure to determine when to fall back to IFF explanations needs to be developed. While these may sound simple to state, domain-specific nuances would entail different domain-specific technical pathways for achieving F\&I. 

\section{Concluding Notes}

We considered the distinctions between fairness and interpretability, and outlined two principles-based frameworks that entail technical challenges where fairness and interpretability can meaningfully work together. We hope that these will contribute to deepening the scholarly debate towards enhancing ethical AI in meaningful ways. 

\bibliographystyle{named}
\bibliography{ijcai20}

\end{document}